\documentclass[preprint]{aastex}
\begin{document}

\title{On the Early In Situ Formation of Pluto's Small Satellites}
\author{Man Yin Woo\altaffilmark{1} and Man Hoi Lee\altaffilmark{1,2}}
\altaffiltext{1}{Department of Earth Sciences, The University of Hong Kong,
  Pokfulam Road, Hong Kong}
  \altaffiltext{2}{Department of Physics, The University of Hong Kong,
  Pokfulam Road, Hong Kong}

\begin{abstract}
The formation of Pluto's small satellites --- Styx, Nix, Keberos and
Hydra --- remains a mystery. Their orbits are nearly circular and are
near mean-motion resonances and nearly coplanar with Charon's
orbit. One scenario suggests that they all formed close to their
current locations from a disk of debris that was ejected from the
Charon-forming impact before the tidal evolution of Charon. The
validity of this scenario is tested by performing $N$-body simulations
with the small satellites treated as test particles and Pluto-Charon
evolving tidally from an initial orbit at a few Pluto radii with
initial eccentricity $e_{\rm C} = 0$ or 0.2. After tidal evolution,
the free eccentricities $e_{\rm free}$ of the test particles are
extracted by applying fast Fourier transformation to the distance
between the test particles and the center of mass of the system and
compared with the current eccentricities of the four small satellites.
The only surviving test particles with $e_{\rm free}$ matching the
eccentricities of the current satellites are those not affected by
mean-motion resonances during the tidal evolution in a model with
Pluto's effective tidal dissipation function $Q = 100$ and an initial
$e_{\rm C}$ = 0.2 that is damped down rapidly. However, these test
particles do not have any preference to be in or near 4:1, 5:1 and 6:1
resonances with Charon. An alternative scenario may be needed to
explain the formation of Pluto's small satellites.
\end{abstract}

\section{INTRODUCTION}

Pluto, the dwarf planet visited by the {\it New Horizons} spacecraft
in July 2015, has  a complex satellite system. With 5 satellites in
total, from closest to furthest, they are Charon, Styx, Nix, Kerberos
and Hydra. Charon, discovered in 1978 (\citealt{chr78}), is the
largest one, which has a radius about half of Pluto's and a mass
around 1/8 of Pluto's (e.g., \citealt{ste15}). The satellite-to-planet
mass ratio is high enough for Pluto and Charon to be regarded as a
binary system since the center of mass of the system is located
outside both bodies. Other objects orbiting around the barycenter of
the binary are the four much smaller satellites that remain
undiscovered until the {\it Hubble Space Telescope} imaged them in the
early 21st century (\citealt{wea06}; \citealt{sho11};
\citealt{sho12}). Table \ref{table0} shows the orbital and physical
parameters of the four small satellites.

Astronomers are intrigued to understand the formation of Pluto's satellites due to their special orbital characteristics. All satellites are orbiting on nearly coplanar and nearly circular orbits (\citealt{bro15}; \citealt{sho15}). The orbital period ratios of the four small satellites are close to 1:3:4:5:6 with respect to Charon. Although close to integer ratios, the period ratios are significantly off from integer ratios, and the four small satellites are not trapped in mean-motion resonances (MMR) with Charon (\citealt{bro15}; \citealt{sho15}). Besides, Nix and Hydra are close to a mutual 3:2 MMR (\citealt{lee06}; \citealt{bro15}; \citealt{sho15}). 

The most widely accepted scenario for the formation of Charon is the
intact capture scenario in which Charon was the impactor captured
in a giant collision that most likely happened when the population of
Kuiper Belt objects was much higher than today (\citealt{can05}).
Charon ended up in an eccentric orbit with semimajor axis about $4
R_{\rm P}$ (where $R_{\rm P} = 1187\,$km is the radius of Pluto)
(\citealt{can05}). Tidal evolution eventually brought them into the
double synchronous state --- the end state of tidal evolution with the
orbital and spin periods of the two bodies being exactly the same and
the orbit having zero eccentricity. Pluto and Charon are now $\sim 17
R_{\rm P}$ from each other.

However, for the small satellites, there is not yet a complete and
consistent scenario to explain their formation and how they ended up
in their unusual orbits.
Several scenarios have been proposed. The first one is known as forced
resonant migration, which was proposed by \cite{war06}. In this
scenario, Nix and Hydra (and presumably Styx and Kerberos, which were
still undiscovered in 2006) were debris from the giant impact that formed
Charon. They formed closer to Pluto than their current locations. When
Charon evolved outward due to tidal evolution, the small satellites
were caught into corotation resonances with Charon and moved outward
together with Charon. The small satellites can be carried to their
current positions without instability, because their orbital
eccentricities would not be forced up by corotation resonances. When
Charon finished tidal evolution, its orbital eccentricity was damped
down to zero and the small satellites would escape from the resonances
with Charon. However, \cite{lit08} ruled out this idea. They found
that in order to transport Nix, Charon's eccentricity $e_{\rm C}$
should be smaller than 0.024. Otherwise, overlapping of the second
order Lindblad resonance with the corotation resonance at 4:1 would
lead to chaos. On the other hand, to transport Hydra, $e_{\rm C}$
should be larger than 0.04. Otherwise, Hydra would slip out of
resonance, because its migration would be faster than libration.
Because these two constraints contradict with each other, forced
resonant migration in corotation resonance is an unsuccessful
scenario.

If transport in corotation resonance does not work, how about
transport by capture into multiple resonances at the same mean-motion
commensurability? \cite{che14b} investigated this situation and found
that although stable capture and transport in 5:1, 6:1, and 7:1
commensurabilities are possible, it is also unlikely for this scenario
to work because their results show that the satellites that survive to
the end of tidal evolution would have significant final
eccentricities. Besides, no stable capture and transport at 3:1 and
4:1 commensurabilities can be observed from their simulations. Because
of the expected initial fast rotation of Pluto, large gravitational
coefficient $J_2$ of Pluto was also tested. They found that large
$J_2$ of Pluto causes the resonances at the same commensurability to
be further apart from each other and hence libration in multiple
resonances simultaneously is much more difficult. This destroys the
condition for stable migration in multiple resonances for the small
satellites.

Another scenario suggested by \cite{lit08}, known as the collisional
capture scenario, was studied by \cite{pir12}. In this scenario, a
planetesimal orbiting around the Sun could be temporarily captured by
the Pluto-Charon binary which finished tidal evolution. The captured
planetesimal collided with another incoming planetesimal and formed a
debris disk. Collisions between bodies in the disk damped down the
orbital eccentricities and inclinations, leading to the formation of a
co-planar and circular debris disk in which the small satellites can
be formed near their current positions. However, this scenario was
ruled out by \cite{pir12}, as they found that the timescale of
temporary capture for objects that are massive enough to produce Nix
and Hydra is much smaller than the timescale for another object to
come in and collide with it. Their assumed masses for Nix and Hydra
are adopted from \cite{tho08}. Besides, \cite{wal15} investigated 
how the debris disk would evolve if such a collision did occur and
found that the satellites formed would have no strong preference to be
in or near resonances. Based on these studies, the collisional capture
scenario may not be the answer.

Similar to the collisional capture scenario, \cite{ken14} suggested
that the four small satellites could be formed near their current
positions but with a different process. In this early in-situ
formation scenario, the Charon forming giant impact produced a debris
ring at around $20 R_{\rm P}$. \cite{ken14} estimated that the
transfer of angular momentum from the central binary to the
ring could lead to the spreading of the ring to the current positions
of the satellites within 5 to 10 years, which is much faster than the
tidal evolution rate of Pluto and Charon. Later, in their second
paper, \cite{brom15} performed a complete study on the spreading of
the collisional disk around the tidally evolving Pluto-Charon. They
found that the spreading of the ring can occur on a timescale
comparable to the formation timescale of the satellites. With such a
rapid spreading process, satellites can form near their current
positions when Charon was still close to Pluto. Recent crater counting
data from {\it New Horizons} imply that the surface ages of Nix and
Hydra are at least 4 billion years  (\citealt{wea16}). This supports
the early formation of the four small satellites. However, the effect
of outward tidal evolution of Charon has not been investigated in this
scenario. The outwardly evolving Charon would perturb the orbits of
the small satellites (for instance, by MMR) and hence they may not be
able to lie on nearly circular and coplanar orbits after Charon
completed its tidal evolution.

In this paper, we test the plausibility of the early in-situ formation
scenario by investigating the effects of the tidal evolution of
Charon's orbit on the small satellites' mean distances and
eccentricities, after the small satellites formed near their current
positions from a debris disk. We also examine the final orbits of the
satellites in order to see if their orbits are near MMR with Charon.
We present our methods and results in Sections \ref{sec:methods} and
\ref{sec:results}, respectively.
The results are summarized and discussed in Section
\ref{sec:conclusions}.

\section{METHODS}
\label{sec:methods}

We adopt the masses and radii of Pluto and Charon from \cite{bro15}:
$GM_{\rm P} = 869.61 \,{\rm km}^3 \, {\rm s}^{-2}$, $GM_{\rm C} =
105.88 \,{\rm km}^3\,{\rm s}^{-2}$, $R_{\rm P} = 1181\,$km, and
$R_{\rm C} = 603.6\,$km, where $M_{\rm P}$, $M_{\rm C}$, $R_{\rm P}$
and $R_{\rm C}$ are the masses and radii of Pluto and Charon and $G$
is the gravitational constant. The gravitational harmonic coefficients
$J_2$ and $C_{22}$ of Pluto and Charon are set to zero throughout our
calculation, where $C_{22}$ is the permanent quadrupole moment.
Based on the giant impact hypothesis, Charon is placed at $a_{\rm PC}
= 4 R_{\rm P}$ initially (\citealt{can05}). The initial orbital eccentricity of Charon, $e_{\rm C}$, is set to be 0 or 0.2. The initial $e_{\rm C} = 0.2$ case corresponds to Charon formation according to the intact capture scenario. Zero eccentricity corresponds to Charon forming from a debris disk generated from the giant impact, since \cite{can05} does not rule out this situation although it is less likely. 
We assume that the initial spin rate of Charon is twice the initial
mean motion. Then the initial spin rate of Pluto is $\sim 5.65$ times
the initial mean motion, calculated from the current total angular
momentum with $a_{\rm PC} = 4 R_{\rm P}$.

For the small satellites, we treat them as massless test particles and
place them randomly in the current orbital distance range of the 4
small satellites, about 35 to $60 R_{\rm P}$ from the barycenter of
the binary. To imitate the situation where the satellites formed in a
collisional debris disk near their current positions, their initial
orbits (coplanar with Charon's orbit) are obtained from an integration
with eccentricity damping applied to the test particles.
We use the Wisdom-Holman \citep{wis91} integrator in the
SWIFT package \citep{lev94}, modified for integrations of systems with
comparable masses such as Pluto-Charon \citep{lee03}.
We apply eccentricity damping to the test particles as half steps
before and after each Wisdom-Holman step \citep{lee02}, with a damping
equation
\begin{equation}
e_{\rm final} = e_{\rm initial}\exp(-0.5k\delta t ),
\label{edamp}
\end{equation}
where $e_{\rm initial}$ and $e_{\rm final}$ are the osculating
eccentricities before and after damping, respectively, $\delta t \sim
630\,$s is the time step we adopted for the damping calculations, and
$k = 10^{-10} \, {\rm s}^{-1}$ is the damping coefficient we
adopted. Eventually, after about 2000 years, the test particles are
damped to nearly the coldest orbits.
For test particle orbiting a binary, the coldest orbit is the one
with the amplitude of the epicyclic motion, i.e., the free
eccentricity $e_{\rm free}$, equal to zero, since the oscillations
forced by the non-axisymmetric components of the binary's potential,
including the forced eccentricity
\begin{equation}
e_{\rm forced}\approx\frac{5}{4}e_{\rm C}\frac{M_{\rm P}-M_{\rm C}}{M_{\rm P}+M_{\rm C}}\frac{a_{\rm PC}}{R_0},
\label{eforce}
\end{equation}
cannot be damped down through collisions between particles within the
debris disk \citep{leu13}.
In Equation (\ref{eforce}),
$a_{\rm PC}$ is the orbital semimajor axis of Pluto-Charon and
$R_0$ is the average distance between the test particle and the
center of mass of Pluto-Charon.
For $e_{\rm C} = 0$, $e_{\rm forced}$ equals to zero according to
Equation (\ref{eforce}). For $e_{\rm C} = 0.2$ and $a_{\rm PC} = 4
R_{\rm P}$, $e_{\rm forced}$ is around 0.01 to 0.02 at the distances
of the test particles.
The free eccentricity $e_{\rm free}$ of the test particles at the end
of the damping calculation is about 10 times smaller than $e_{\rm
  forced}$.

\begin{figure}
\begin{center}
\includegraphics[width=0.8\textwidth]{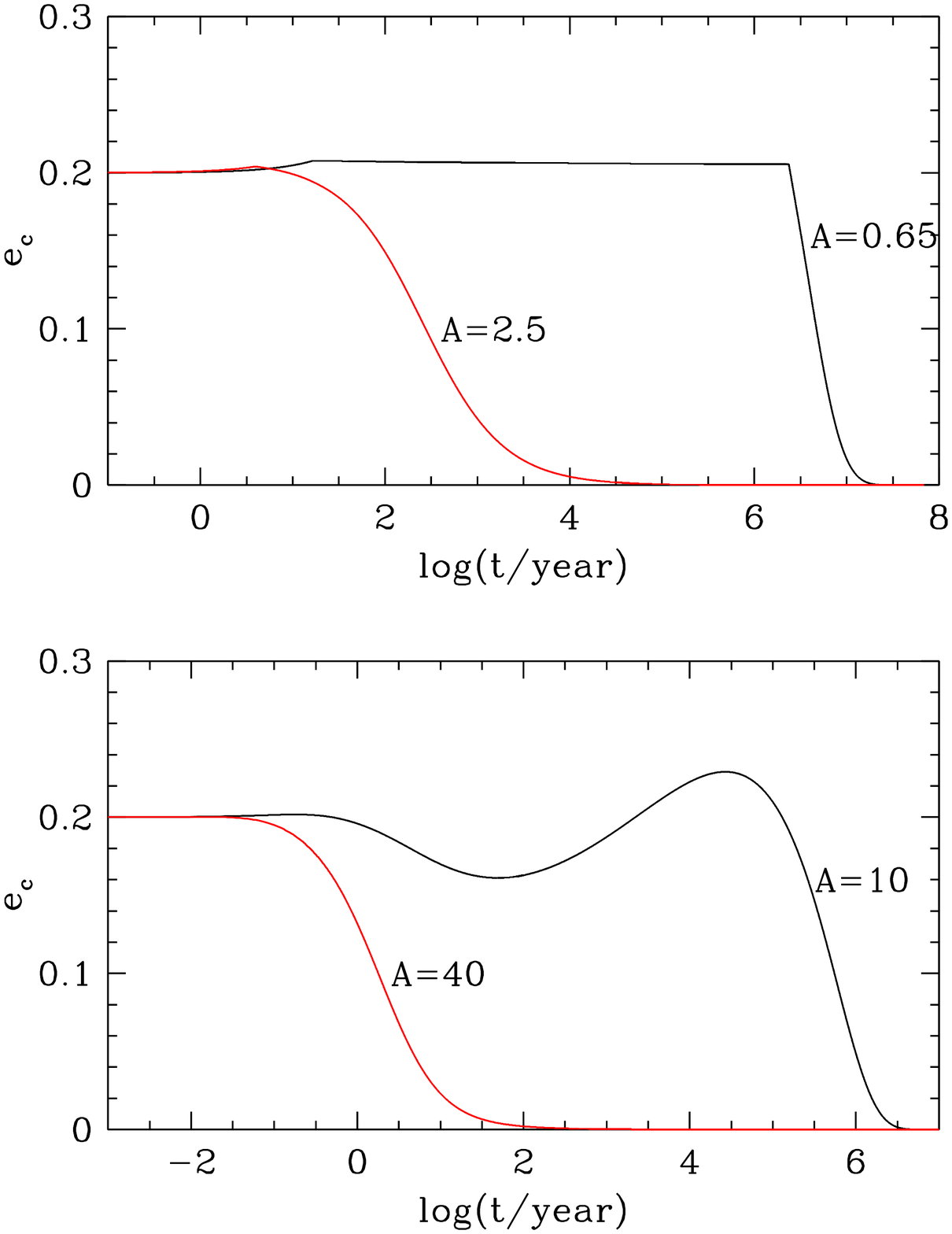}
\caption{
Tidal evolution of the orbital eccentricity of Charon, $e_{\rm C}$.
The upper panel shows the constant $Q$ model with $Q = 100$, and the lower panel shows the
constant $\Delta t$ model with $\Delta t = 600\,$s. In each panel, the black line is the integration with smaller $A$ value, and the red line is the integration with larger $A$ value.
\label{fig1}
}
\end{center}
\end{figure}

After the eccentricity damping calculation, the whole system is then
integrated in the tidal evolution code developed by \cite{che14a}. Two
tidal models are used: constant $\Delta t$ and constant $Q$, where
$\Delta t$ is the time lag of the tidal bulge of Pluto and $Q$ is the
effective tidal dissipation function of Pluto. In the constant $\Delta
t$ model, the phase angle of Pluto's tidal bulge $\alpha =
\sigma_{lm}\Delta t$ when $\alpha$ is small, where $\sigma_{lm}$ is
the tidal frequency which depends on the mean motion and spin angular
velocity of Pluto. Thus $\alpha$ changes continuously throughout the
tidal evolution. In the constant $Q$ model, $\alpha = {\rm
  sgn}(\sigma_{lm})/Q$ when $\alpha$ is small, and $\alpha$ is
independent of the tidal frequency, except for the sign. Since
$\alpha$ has different dependence on tidal frequency, orbital
evolution in the constant $Q$ model is qualitatively different from
the evolution in the constant $\Delta t$ model, especially near the
end of tidal evolution (see Figure \ref{fig1}).

Another major difference between the two models is the evolution
timescale. We adopt $\Delta t = 600\,$s for constant $\Delta t$ and
$Q = 100$ for constant $Q$, as in \cite{che14a}. Due to the values of
$\Delta t$ and $Q$ adopted, constant $Q$ models would take 5 to 10
times longer to evolve to the doubly synchronous state than the
constant $\Delta t$ models. We adopt an initial integration time step
of $1000\,$s.
For constant $\Delta t$, due to the expansion of the orbital semimajor
axis of Charon, we increase the time step to $5000\,$s when Charon
evolves to $\sim 10 R_{\rm P}$ from Pluto, until the end of tidal
evolution ($\sim 10^6$ years). For constant $Q$, we adopt a similar
procedure, except for an extra increase of the time step to $10000\,$s
at $\sim 3.17 \times 10^6$ years after the start of the simulation,
due to the longer tidal evolution time ($\sim 10^7$ years).
The readers are referred to \cite{che14a} for details of the tidal
models and the reasons for adopting various parameter values for the
tidal models and the numerical integrations.

Two values for the relative rate of tidal dissipation in Charon and
Pluto, $A$, are adopted for each model. For constant $\Delta t$,  
\begin{equation}
A = \frac{k_{2C}}{k_{2P}} \frac{\Delta{t_{\rm C}}}{\Delta{t}}\left ( \frac{M_{\rm P}}{M_{\rm C}} \right )^2\left ( \frac{R_{\rm C}}{R_{\rm P}} \right )^5,
\end{equation}
where $k_{2C}$ and $k_{2P}$ are the Love numbers of Charon and Pluto, respectively, and $\Delta t_{\rm C}$ is the time lag of Charon's tidal bulge \citep{mig80,che14a}. For constant $Q$, 
\begin{equation}
A = \frac{k_{2C}}{k_{2P}} \frac{Q}{Q_{\rm C}}\left ( \frac{M_{\rm P}}{M_{\rm C}} \right )^2\left ( \frac{R_{\rm C}}{R_{\rm P}} \right )^5,
\end{equation}
where $Q_{\rm C}$ is the effective tidal dissipation function of Charon \citep{yod81,che14a}. We adopt $A = 10$ or 40 for constant $\Delta t$ and $A = 0.65$ or 2.5 for constant $Q$. 
Figure \ref{fig1} shows the evolution of Charon's eccentricity
throughout the tidal evolution in the constant $\Delta t$ and constant
$Q$ models, with initial $e_{\rm C} = 0.2$. For the smaller $A$ values
($A = 10$ for constant $\Delta t$ and $A = 0.65$ for constant $Q$),
$e_{\rm C}$ stays around 0.2 until near the end of tidal
evolution. For the larger $A$ values ($A = 40$ for constant $\Delta t$ and $A = 2.5$ for constant $Q$), $e_{\rm C}$ damps down rapidly to near zero when $t \sim 10^3$ years for constant $\Delta t$ and $t \sim 10^5$ years for constant $Q$. Therefore, the damping rate of $e_{\rm C}$ depends on the value of $A$. 

We integrate 200 test particles with different initial semimajor axes
and eccentricities (before the eccentricity damping calculation) for
each combination of $A$ and initial $e_{\rm C}$. We should emphasize
that we are not modeling a debris disk but a set of individual satellites that have already formed within a debris disk. Table \ref{table1} shows the combination of $A$ and $e_{\rm C}$ we integrate in each tidal model. For initial $e_{\rm C} = 0$, we only integrate with larger $A$, since $e_{\rm C}$ would stay at zero throughout the tidal evolution and different $A$ values would not affect the evolution of $e_{\rm C}$.

\begin{figure}
\begin{center}
\includegraphics[width=0.8\textwidth]{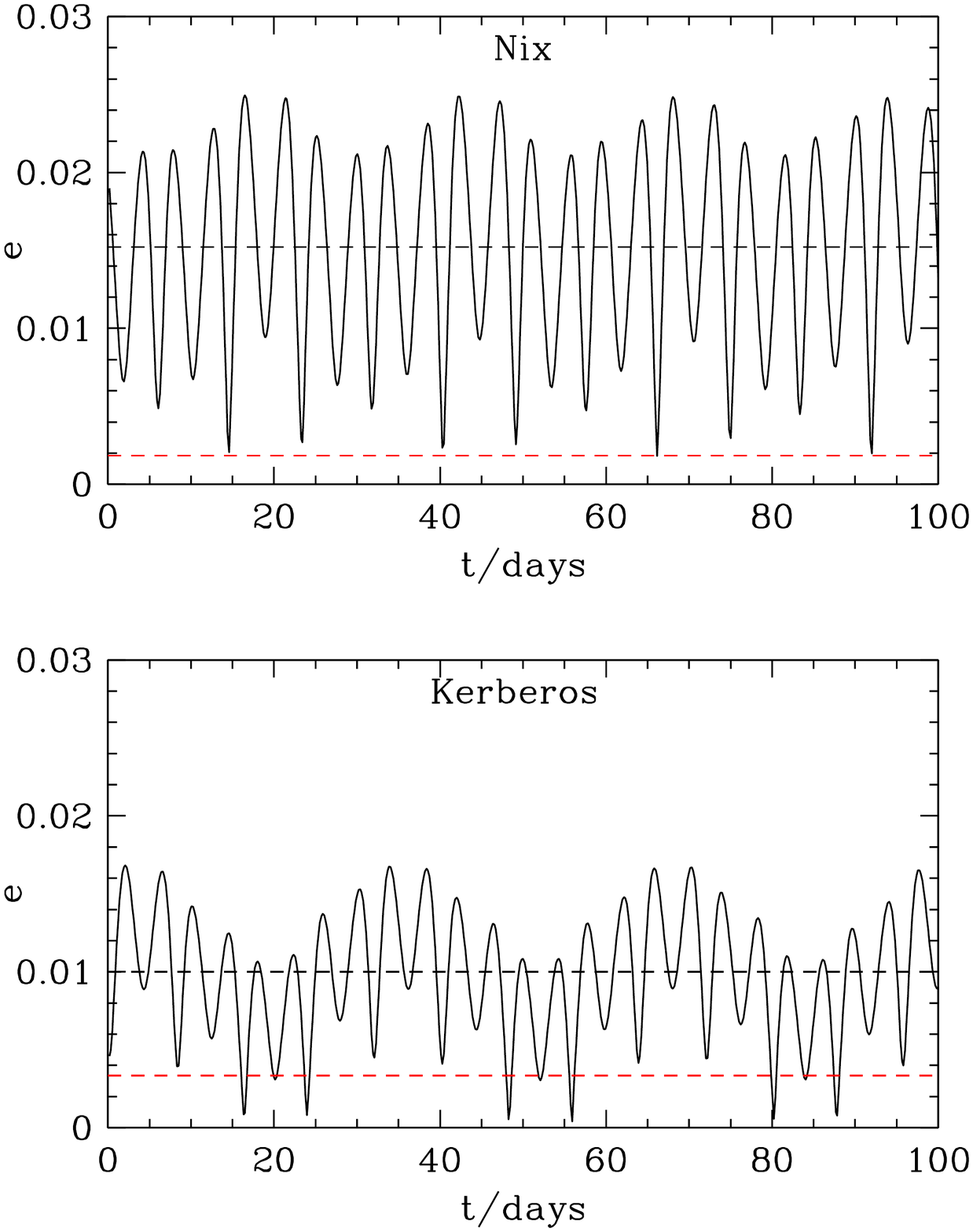}
\caption{
Osculating eccentricity evolution of Nix and Kerberos from the best fit data of \cite{sho15}. The black solid and dashed lines are the osculating eccentricity $e_{\rm osc}$ and the mean of $e_{\rm osc}$, respectively. The red dashed lines are the free eccentricities $e_{\rm free}$ obtained from FFT. 
\label{fig2}
}
\end{center}
\end{figure}

After the tidal evolution of Charon, the surviving test particles are
then integrated for another 800 days and we apply fast Fourier
transformation (FFT) to the distance $R(t)$ between the test particle
and the center of mass of the system to calculate the magnitude of the
free eccentricity, $e_{\rm free}$, from the power in the peak at the
epicyclic frequency $\kappa_0$ in the power spectrum (see
\citealt{woo17} for details). The reason for extracting $e_{\rm free}$
of the test particles from the power spectrum obtained from FFT is
that the osculating Keplerian eccentricity $e_{\rm osc}$ can show
significant variations due to the oscillations forced by the
Pluto-Charon binary and that even the mean of $e_{\rm osc}$ can be
significantly different from $e_{\rm free}$. In addition, $e_{\rm 
  free}$ of the current satellites are closer to the eccentricities
obtained by \cite{sho15} from fitting the orbits of the small
satellites by precessing ellipses.
Figure \ref{fig2} shows the orbital integration of two of the small
satellites, Nix and Kerberos, using the best fit data of \cite{sho15}.
We observe that there are obvious differences between $e_{\rm osc}$
and $e_{\rm free}$ for each satellite, with the mean of $e_{\rm osc}$
at least a factor of 3 larger than $e_{\rm free}$, and that
$e_{\rm free}$ is much closer to the fitting result of \cite{sho15}
listed in Table \ref{table0}.
Therefore, we decide to compare $e_{\rm free}$ of the test particles
with the eccentricities of the small satellites listed in Table
\ref{table0}.

The adopted initial integration time step of $1000\,$s is more than 60
steps per orbit of Pluto-Charon for initial $a_{\rm PC} = 4 R_{\rm P}$
and resolves the dominant forced oscillations in the motion of the
test particles.
We repeat one set of simulations (constant $\Delta t$ with $A = 10$
and initial $e_{\rm C} = 0.2$) with half and twice the adopted time
step.
We find that the statistics are identical to those in Table
\ref{table2} within uncertainties, which confirm that the adopted time
step is small enough.

\section{RESULTS}
\label{sec:results}

We first define the resonant terms in the disturbing function in order to explain the resonant behaviors of the test particles. For coplanar orbits, the lowest order resonant terms at the $m + 1$:1 mean-motion commensurability exterior to Charon are 
\begin{equation}
\Phi_m = {GM_{\rm C}\over a}\sum_{n = 0}^mf_{m,n}(\alpha)e^{m-n}e_{\rm C}^n\cos\phi_{m,n},
\label{res_term}
\end{equation}
where $\phi_{m,n} = (m + 1) \lambda - \lambda_{\rm C} - (m - n)\varpi - n\varpi_{\rm C}$ are the resonant angles, $a$, $e$, $\lambda$, and $\varpi$ are the orbital semimajor axis, eccentricity, mean longitude, and longitude of periapse of the small satellite, and the orbital elements with subscript C are those of Charon \citep{mur99}. $f_{m,n}$ are functions of $\alpha = a_{\rm C}/a$, the Laplace coefficients and their derivatives with respect to $\alpha$.

\begin{figure}
\begin{center}
\includegraphics[angle=270,width=1.1\textwidth]{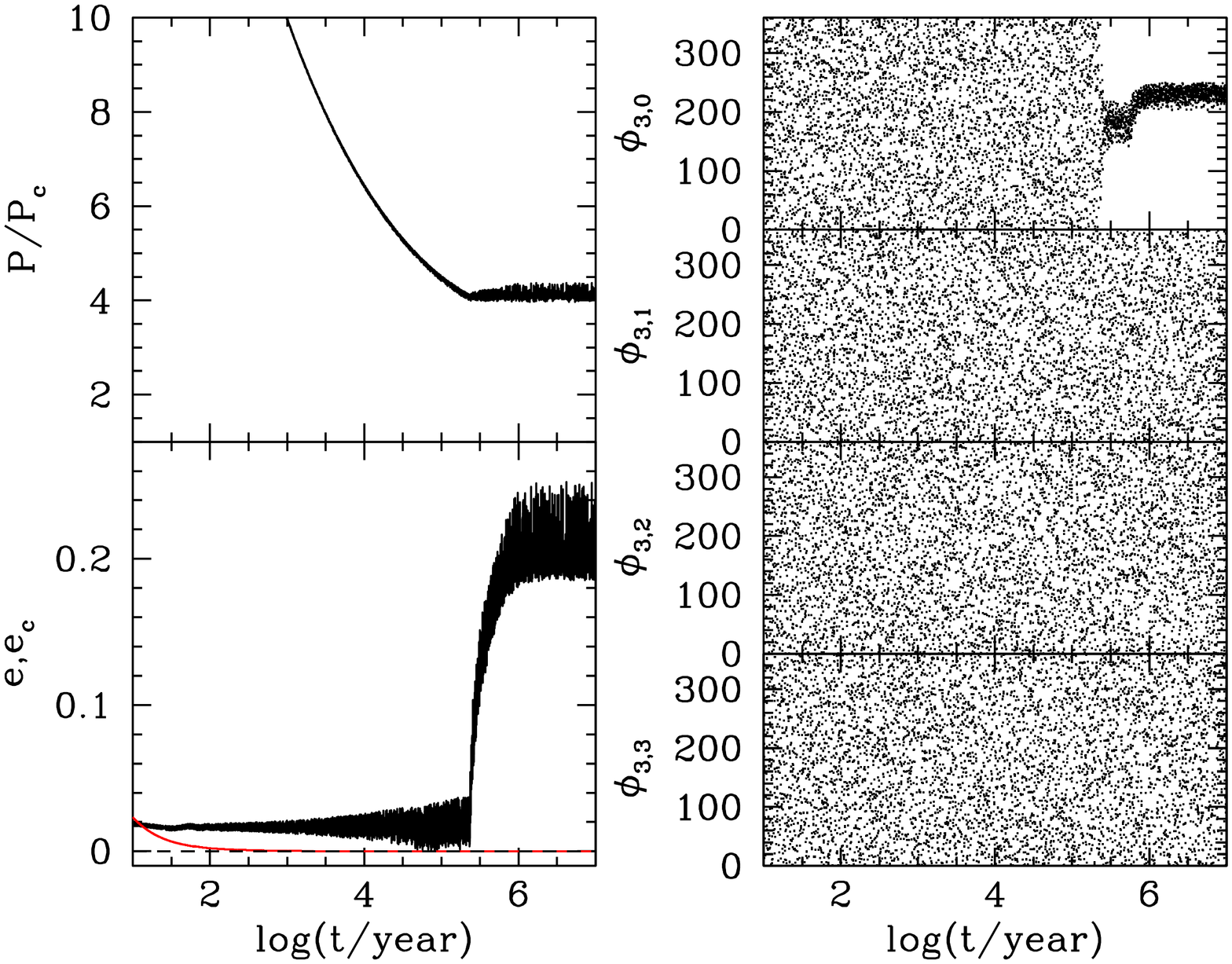}
\caption{
Evolution of the period ratio with respect to Charon (upper left
panel), $e_{\rm osc}$ (lower left panel), and the four resonant angles
$\phi_{3,n}  = 4\lambda - \lambda_{\rm C} - (3 - n)\varpi -
n\varpi_{\rm C}$ of the 4:1 resonance (right panels) for a test
particle that is still trapped in resonance at the end of tidal
evolution, in the constant $\Delta t$ model with $A = 40$ and initial
$e_{\rm C} = 0.2$.
The red line in the lower left panel shows the evolution of $e_{\rm C}$ (see Figure \ref{fig1} for the full evolution of $e_{\rm C}$).
\label{fig3}
}
\end{center}
\end{figure}

\begin{figure}
\begin{center}
\includegraphics[angle=270,width=1.1\textwidth]{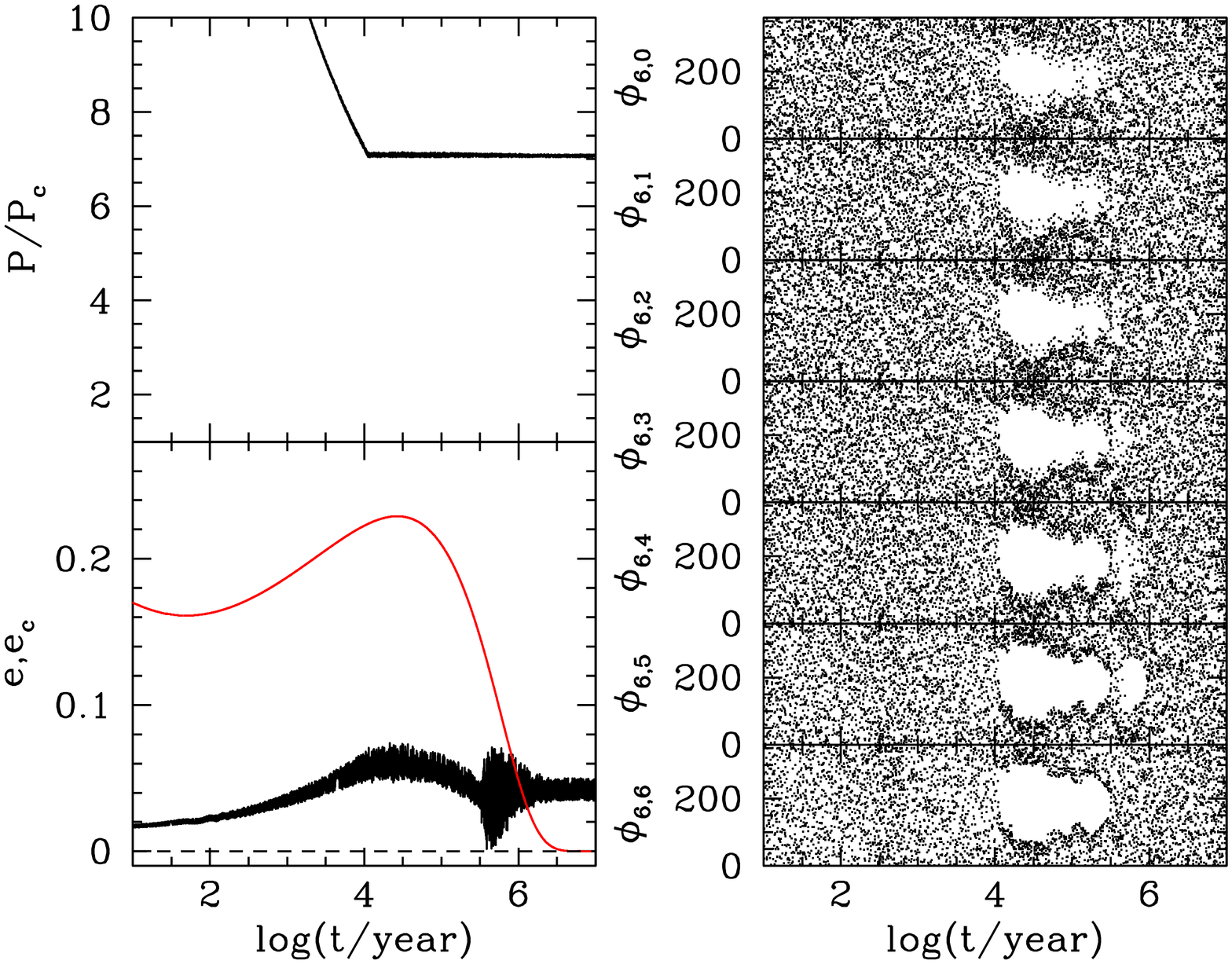}
\caption{
Evolution of the period ratio with respect to Charon (upper left
panel), $e_{\rm osc}$ (lower left panel), and the seven resonant angles
$\phi_{6,n}  = 7\lambda - \lambda_{\rm C} - (6 - n)\varpi -
n\varpi_{\rm C}$ of the 7:1 resonance (right panels) for a test
particle that is trapped but then escape from resonance, in the
constant $\Delta t$ model with $A = 10$ and initial $e_{\rm C} = 0.2$.
The red line in the lower left panel shows the evolution of $e_{\rm C}$.
\label{fig4}
}
\end{center}
\end{figure}

\begin{figure}
\begin{center}
\includegraphics[angle=270,width=1.1\textwidth]{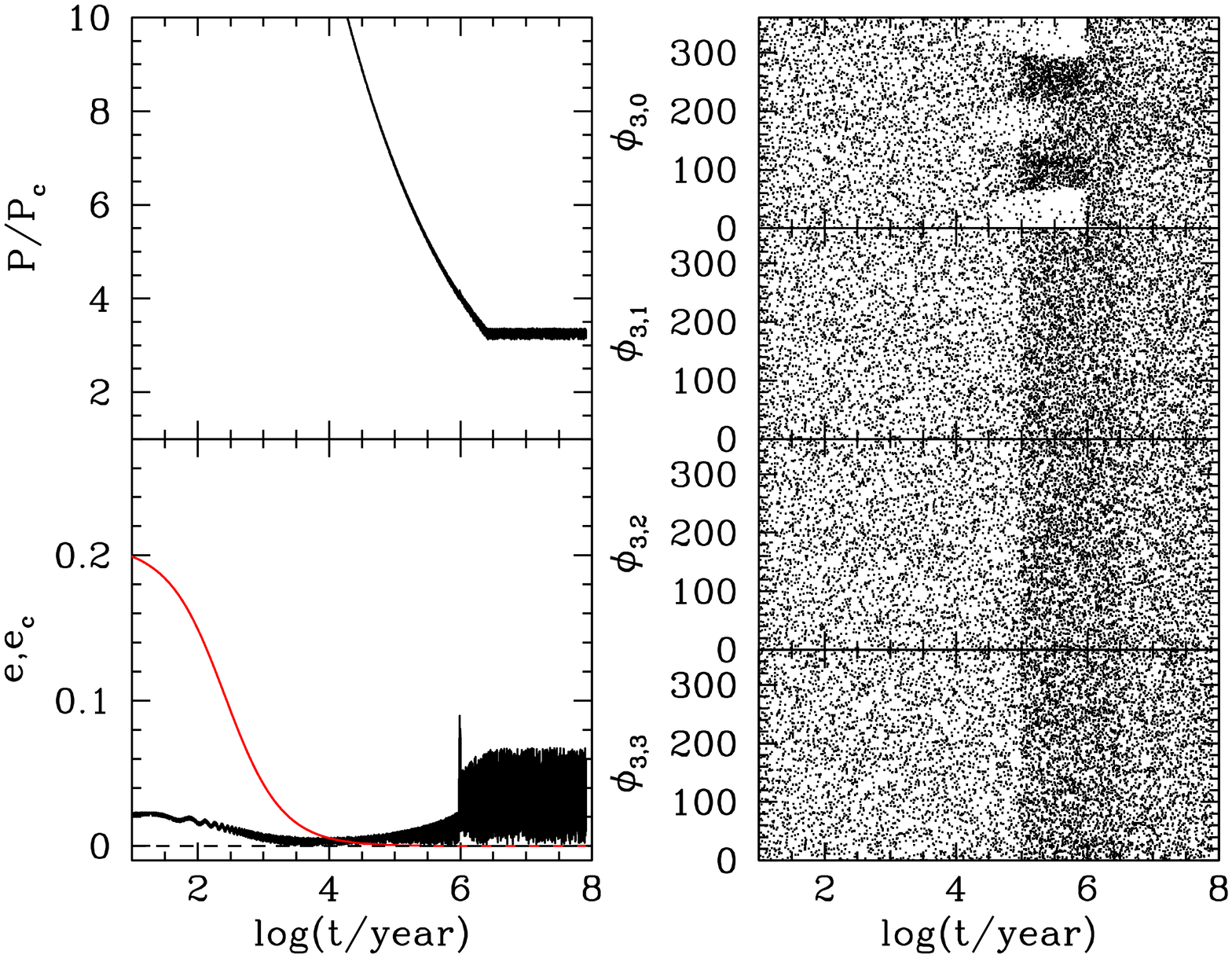}
\caption{
Same as Figure \ref{fig3}, but for a test particle that passes through
the 4:1 MMR and is affected but not trapped in the 4:1 MMR, in the constant $Q$ model with $A =2.5$ and initial $e_{\rm C} = 0.2$. The change in density of points in the right panels after $t = 10^5$ years is due to a change in data sampling frequency.
\label{fig5}
}
\end{center}
\end{figure}

\begin{figure}
\begin{center}
\includegraphics[angle=270,width=1.1\textwidth]{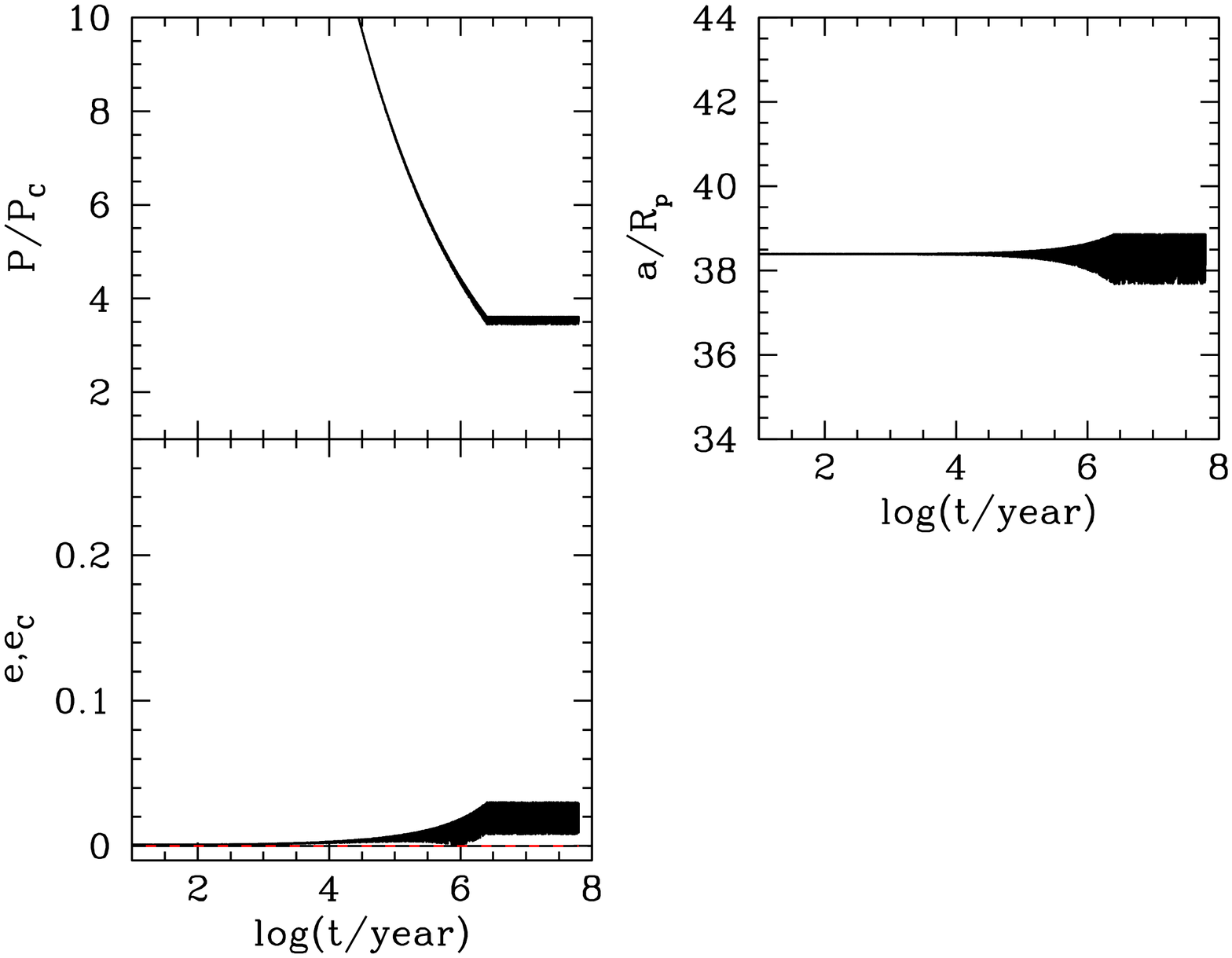}
\caption{
Evolution of the period ratio with respect to Charon (upper left panel), $e_{\rm osc}$ (lower left panel), and the osculating semimajor axis $a_{\rm osc}$ (upper right panel) for a test particle that is not affected by resonances, in the constant $Q$ model with $A = 2.5$ and initial $e_{\rm C} = 0$. The red line in the lower left panel shows the evolution of $e_{\rm C}$ which stays at zero throughout the whole tidal evolution.
\label{fig6}
}
\end{center}
\end{figure}

We now present the results of our simulations. Some of the test particles are ejected during the tidal evolution due to trapping in MMR with Charon. Their eccentricities are forced up to extremely high values. The evolution of the test particles that survive to the end of the tidal evolution falls into 4 categories : 

\begin{enumerate}

\item Some test particles are still trapped in MMR
  with Charon at the end of the tidal evolution (labeled as ``Still
  trapped in MMR" in Table \ref{table2}). Figure \ref{fig3} shows a test particle that is
  trapped in 4:1 MMR with Charon, in the constant $\Delta t$ model
  with $A = 40$ and initial $e_{\rm C} = 0.2$. One of the
  resonant angles $\phi_{3,0}  = 4\lambda - \lambda_{\rm C} -
  3\varpi$ is librating around $\sim 220^{\circ}$ at the end of
  tidal evolution. The reason is that $e_{\rm C}$ is
  damped down to zero when the test particle is trapped into 4:1
  MMR. Hence, only the resonant term in $\Phi_3$ with $\phi_{3,0}$,
  which does not depends on $e_{\rm C}$, is effective at the 4:1
  commensurability (see Equation (\ref{res_term})).  Also, $P/P_{\rm
    C}$ ends near 4 and the eccentricity of the test particle is
  forced up to $\sim 0.2$, which is too large compared to the current
  eccentricities of the small satellites. Test particles in this
  category are shown as blue points in Figures \ref{fig7}, \ref{fig8},
  and \ref{fig9}.

\item Some test particles are once trapped in MMR but then escape
  from the resonance when Charon finishes its tidal evolution (labeled as
  ``Once trapped in MMR" in Table \ref{table2}). Figure \ref{fig4} is an example of a test particle
  that is once trapped in the 7:1 MMR with Charon and then escapes from
  it, in the constant $\Delta t$ model with $A = 10$ and initial $e_{\rm C} =
  0.2$. We observe that between $\sim 10^4$ and $\sim 10^6$ years all 7
  resonant angles, $\phi_{6,n}  = 7\lambda - \lambda_{\rm C}
  - (6 - n)\varpi - n\varpi_{\rm C}$ where $n$ are integers from 0 to
  6, are librating. Although $P/P_{\rm C}$ ends near 7, all
    resonant angles are no longer librating but are circulating after
    $10^6$ years, which indicates that the test particle escapes from
    the 7:1 MMR before the end of the tidal evolution. One of the possible
    reasons for the escape from MMR is the decreasing
    $e_{\rm C}$ from $\sim 10^{5}$ years to the end of the tidal
    evolution. This would weaken all of the resonant terms in $\Phi_6$
    (except the one with $\phi_{6,0}$). Test particles in this
    category are shown as red points in Figures \ref{fig7} and \ref{fig8}.

\item Some test particles are perturbed but not trapped in MMR, and the
  eccentricities of the test particles are significantly affected (labeled
  as ``Affected by MMR" in Table \ref{table2}). Figure \ref{fig5} is an example of a test particle
  that is affected by the 4:1 MMR with Charon when it passes through that
  resonance, in the constant $Q$ model with $A = 2.5$ and initial $e_{\rm C} =
  0.2$. The osculating eccentricity $e_{\rm osc}$ of the test particle
  is forced up to $\sim 0.08$ when the test particle is passing
  through the 4:1 commensurability at $\sim 10^6$ years but none of
  the resonant angles, $\phi_{3,n}  = 4\lambda - \lambda_{\rm C} - (3
  - n)\varpi - n\varpi_{\rm C}$ where $n$ are integers from 0 to 3, librate.
  Although the test particle tries to get into $\phi_{3,0}$ in between
  $10^5$ and $10^6$ years, but $\phi_{3,0}$ fails to librate.  This
  shows that the test particle is not trapped into the 4:1 MMR with Charon. Test particles in this category are shown as black points in Figures \ref{fig7}, \ref{fig8}, and \ref{fig9}. 

\item Some test particles are not affected by any MMR when they pass through them (labeled as ``Unaffected by MMR" in Table \ref{table2}). 
Figure \ref{fig6} demonstrates a case where the test particle is not
affected by MMR, in the constant $Q$ model with $A$ = 2.5 and initial $e_{\rm C}
= 0$. The test particle's final $e_{\rm osc}$ remains at a relatively low
value ($\sim 0.02$) and the final period ratio is not near any
integer value. The increase in fluctuations of $e_{\rm osc}$ and the
osculating semimajor axis $a_{\rm osc}$ is due to the orbit of the
test particle becoming less Keplerian. The tidal expansion of the
orbit of Charon increases the effects of the forced oscillation terms.
Test particles in this category are shown as green points in Figures
\ref{fig8} and \ref{fig9}.

\end{enumerate}

Table \ref{table2} shows the statistics of the test particles for the 6 models, which can be grouped into 3 categories (see Table \ref{table1} for the combinations of $A$ and initial $e_{\rm C}$ we adopted):
(1) small $A$ and initial $e_{\rm C} = 0.2$, where $A = 10$ for
  constant $\Delta t$ and $A = 0.65$ for constant $Q$;
(2) large $A$ and initial $e_{\rm C} = 0.2$, where $A = 40$ for
  constant $\Delta t$ and $A = 2.5$ for constant $Q$; and
(3) large $A$ and initial $e_{\rm C} = 0$, where $A = 40$ for
  constant $\Delta t$ and $A = 2.5$ for constant $Q$.

\subsection{Small $A$ and Initial $e_{\rm C} = 0.2$}

Of the $2 \times 200$ test particles (we integrate the same set of 200
test particles in the constant $\Delta t$ and constant $Q$ models),
most of them are ejected by the tidally evolving Charon (see Table
\ref{table2}). Because the orbit of Charon stays at high eccentricity
until near the end of tidal evolution (see Figure \ref{fig1}), the high-order resonant terms in the disturbing function (e.g., $\Phi_5$ or $\Phi_6$ in Equation (\ref{res_term})) are strong enough to allow most of the test particles to be trapped into a high-order MMR with Charon (e.g., 6:1 and 7:1) in the early stages of the tidal evolution. This would force up the eccentricities of test particles \citep{war06} and the test particles are ejected from their orbits eventually.

For the remaining $\sim 12\%$ of the test particles that survive to
the end of tidal evolution, all are affected by resonances. This is
again due to the large $e_{\rm C}$ causing the high-order resonant
terms in the disturbing function to be strong. Among the surviving
test particles, most of them ($\sim 70\%$) are affected by passing
through resonances but are not trapped in resonances. Only $\sim 20\%$
are once trapped in resonances but then
escape from resonances, and $\sim 10\%$ are still trapped in resonances
at the end of tidal evolution. There are some differences between the
results for constant $\Delta t$ and constant $Q$. For example, $\sim
95\%$ of the test particles in the constant $Q$ model are ejected,
compared to only $\sim 80\%$ in the constant $\Delta t$ model. Also,
among the surviving test particles in the constant $\Delta t$ model,
75\% are only affected but not trapped in MMR, whereas only 50\% of
the surviving test particles in the constant $Q$ model are in this
category. The difference in resonance behavior between constant
$\Delta t$ and constant $Q$ could be due to the difference in tidal
evolution timescale. Since Charon evolves much slower in the constant
$Q$ model, it would be more probable for the test particles to be trapped in MMR with Charon in constant $Q$ than in constant $\Delta t$. 

\begin{figure}
\begin{center}
\includegraphics[angle=270,width=1.1\textwidth]{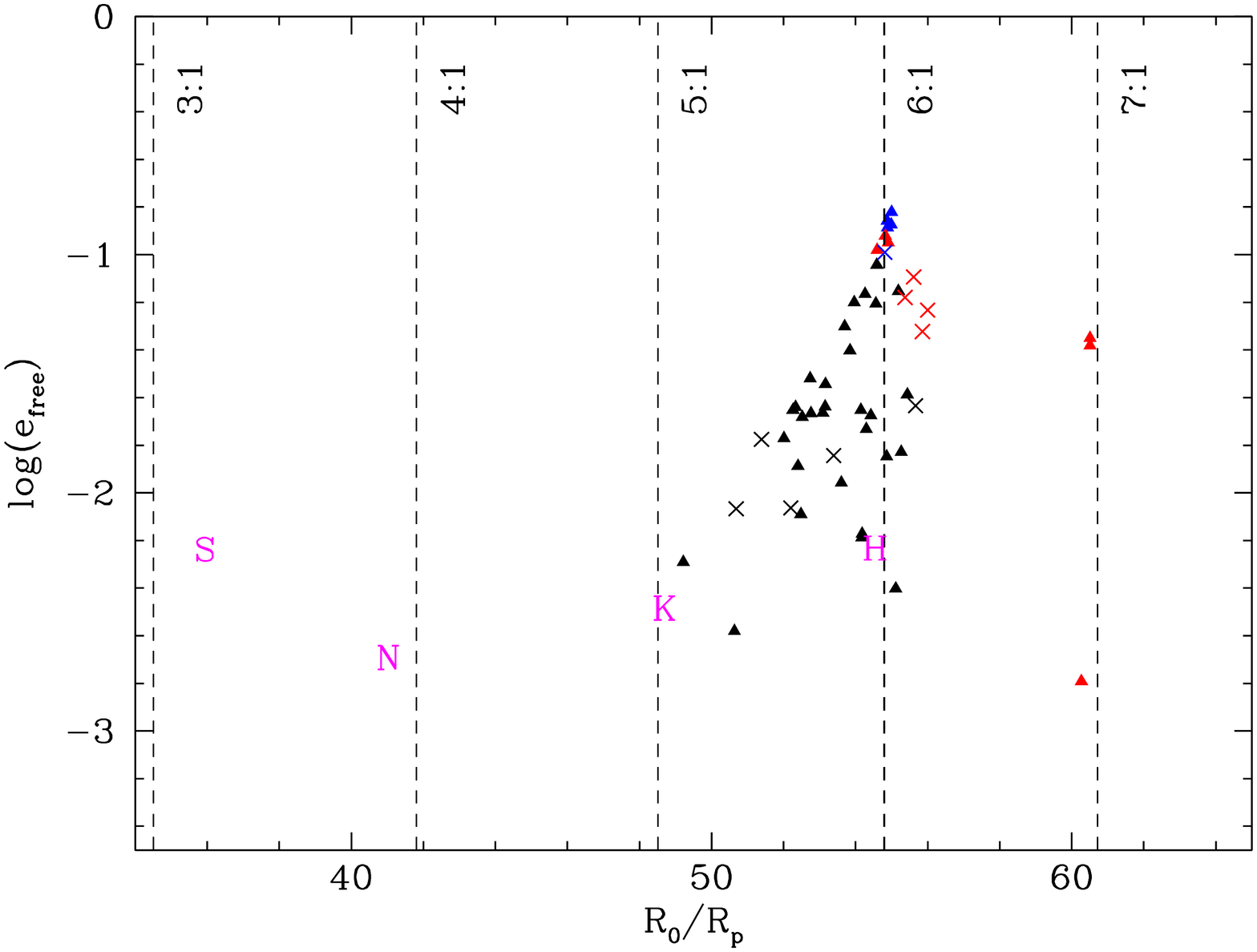}
\caption{
Plot of final log($e_{\rm free}$) against final mean distance to the
center of mass of the Pluto-Charon binary, $R_0$, of the surviving
test particles in the small $A$ and initial $e_{\rm C} = 0.2$
integrations. The blue points are test particles that are still
trapped in resonance at the end of tidal evolution; the red points are
test particles that are trapped but then escape from resonance; and
the black points are the test particles that pass through and are
affected by resonances but are not trapped. The triangles and crosses
correspond to the constant $\Delta t$ and constant $Q$ models,
respectively. The four magenta letters indicate $a$ and $\log(e)$ of
the current four satellites from \cite{sho15} (see
Table \ref{table0}). The current 3:1 to 7:1 mean-motion commensurabilities with Charon are shown by the vertical dashed lines.
\label{fig7}
}
\end{center}
\end{figure}

Figure \ref{fig7} shows the logarithm of $e_{\rm free}$ of the
surviving test particles against their final mean distance to the
center of mass of the Pluto-Charon binary, $R_0$. Compared to the
current 4 satellites (magenta capital letters in Figure \ref{fig7}),
we find some test particles with orbits close to those of Kerberos and
Hydra (``K" and ``H" in Figure \ref{fig7}). However, nothing matches
the orbits of Nix and Styx, since all surviving test particles are
located further than the 5:1 MMR with Charon but Nix and Styx are
located closer than the 4:1 MMR. Hence, this scenario is unable to
explain the formation of Styx and Nix.

Since the number of surviving test particles (40 for constant $\Delta
t$ and 10 for constant $Q$) is quite small in our simulations with
$200$ test particles each, there is a possibility that we have missed
some survivors that match the orbits of Nix and Styx with a reasonable
probability amongst the survivors.
However, we can rule this out for two reasons.
First, we perform an additional set of 200 test particles for each
model and the statistics are consistent with those listed in Table
\ref{table2} and shown in Figure \ref{fig7}, with no surviving test
particles located closer than the 5:1 MMR.
Second, as pointed out by \cite{smu17}, the lack of surviving test
particles closer than the 5:1 MMR can be explained by applying the
instability boundary of \cite{hol99} for circumbinary orbits to the
tidally evolving Pluto-Charon binary.
They found that the instability boundary evolves beyond the orbits of
Styx and Nix before it shrinks back to just inside the orbit of Styx,
if $e_{\rm C}$ remains high until near the end of tidal evolution.

\subsection{Large $A$ and Initial $e_{\rm C} = 0.2$}

Compared to the small $A$ case (where $\sim 88\%$ of the test
particles are ejected), far fewer test particles ($\sim 24\%$) are
ejected (see
Table \ref{table2}). This is because $e_{\rm C}$ damps down to 0 very
quickly when $A$ is large (see Figure \ref{fig1}). When the regions of
high-order resonances like 6:1 and 7:1 start to pass through the
orbits of the test particles, $e_{\rm C}$ already damps down to nearly
0. Only terms with $n = 0$ in Equation (\ref{res_term}) are effective
and the strength of each term mainly depends on the order of the
resonance, $m$. The resonant term is stronger if $m$ is smaller.
Hence, the test particles are less likely to be trapped in
high-order resonances and have a higher chance of survival, unless some
lower order resonances (e.g., 4:1 or 5:1 resonance) sweep through
their orbits. For the same reason, most of the surviving test
particles ($\sim 73\%$) are not affected by resonances.  

\begin{figure}
\begin{center}
\includegraphics[angle=270,width=1.1\textwidth]{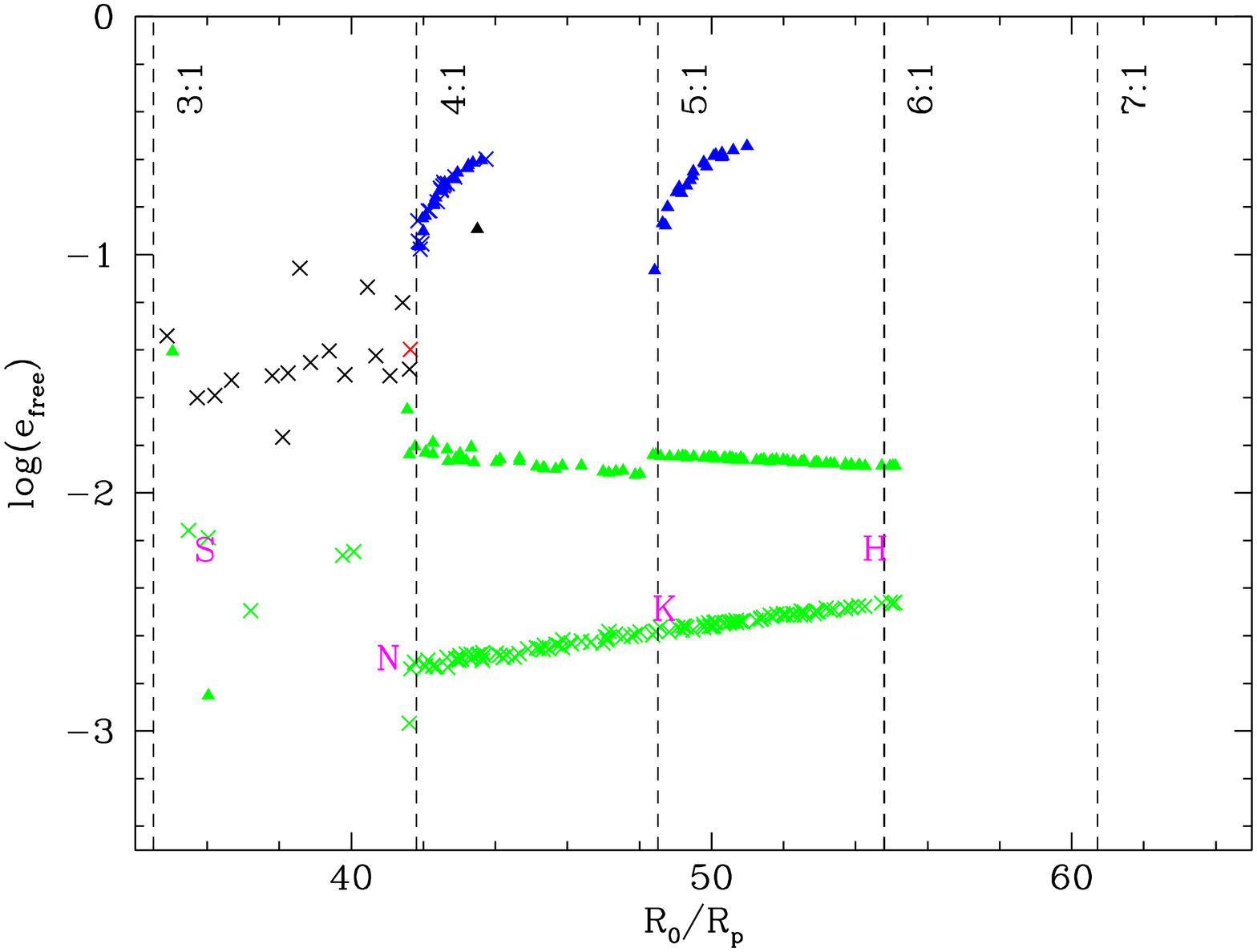}
\caption{
Same as Figure \ref{fig7}, but for the large $A$ and initial $e_{\rm C} = 0.2$ integrations. The green points are test particles that are not affected by resonances.
\label{fig8}
}
\end{center}
\end{figure}

Figure \ref{fig8} shows the plot of final log($e_{\rm free}$) against
final $R_0$ for the test particles that survive to the end of tidal
evolution for large $A$ and initial $e_{\rm C} = 0.2$. We observe that
the blue points ($\sim 21\%$ of the surviving test particles), which
represent test particles that are still trapped in MMR with Charon at
the end of tidal evolution, are only located close to 4:1 and 5:1 MMR.
None is found near the 3:1, 6:1 or 7:1 MMR. On the
other hand, most of the test particles that pass through and are affected
by resonances (black points in Figure \ref{fig8}, which are $\sim 6\%$ of the surviving test particles) end in between the 3:1 and 4:1 MMR. 

For the test particles not affected by resonances (green points in
Figure \ref{fig8}), we find that the results are different for the
constant $\Delta t$ and constant $Q$ models. For constant $\Delta t$,
most of the test particles have $e_{\rm free}$ (green triangles in
Figure \ref{fig8}) within the range of 0.01 to 0.03, whereas most of
the test particles for constant $Q$ have $e_{\rm free}$ (green crosses
in Figure \ref{fig8}) within the range of $10^{-3}$ to $4 \times 10^{-3}$,
which is an order of magnitude lower than those for constant $\Delta
t$. We also observe that in between the 4:1 and 6:1 MMR, the green crosses
form a line with a positive slope (i.e., $e_{\rm free}$ is larger when
the test particle is located further from Pluto-Charon), whereas the
green triangles almost form a line with a negative slope (i.e.,
$e_{\rm free}$ is smaller when the test particle is located further
from Pluto-Charon), except for a jump at the 5:1 MMR.

Compared to the eccentricities of the current satellites, we discover
that some of the test particles in the constant $Q$ model that are not
affected by resonances (green crosses in Figure \ref{fig8}) have
$e_{\rm free}$ that match the current four satellites, whereas no test
particles match the current four satellites in the constant $\Delta t$
model.
The probability of having test particles that are not affected by
resonances is high (over 60\%) in the constant $Q$ model. Although the
test particles that are not affected by resonances in the constant $Q$
model replicate the trend of Nix, Kerberos and Hydra in Figure
\ref{fig8}, no preference for near resonant locations (in between 4:1
and 6:1 MMR) can be observed.

\begin{figure}
\begin{center}
\includegraphics[angle=270,width=1.1\textwidth]{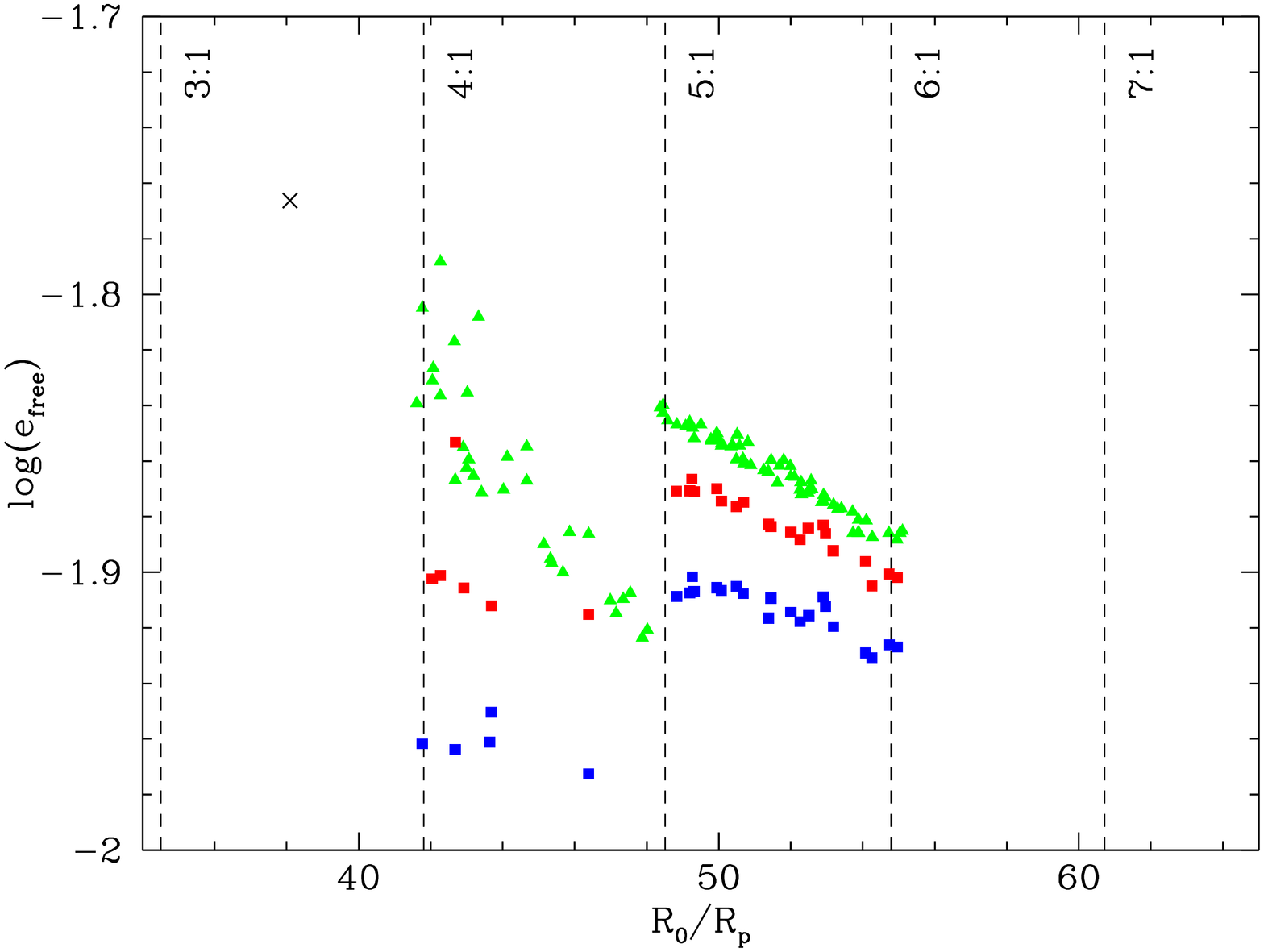}
\caption{
Same as Figure \ref{fig8}, but also with test particles that survive
in the integrations with $\Delta t = 300\,$s (red squares) and
$150\,$s (blue squares).
\label{fig8.1}
}
\end{center}
\end{figure}

\begin{figure}
\begin{center}
\includegraphics[angle=270,width=1.1\textwidth]{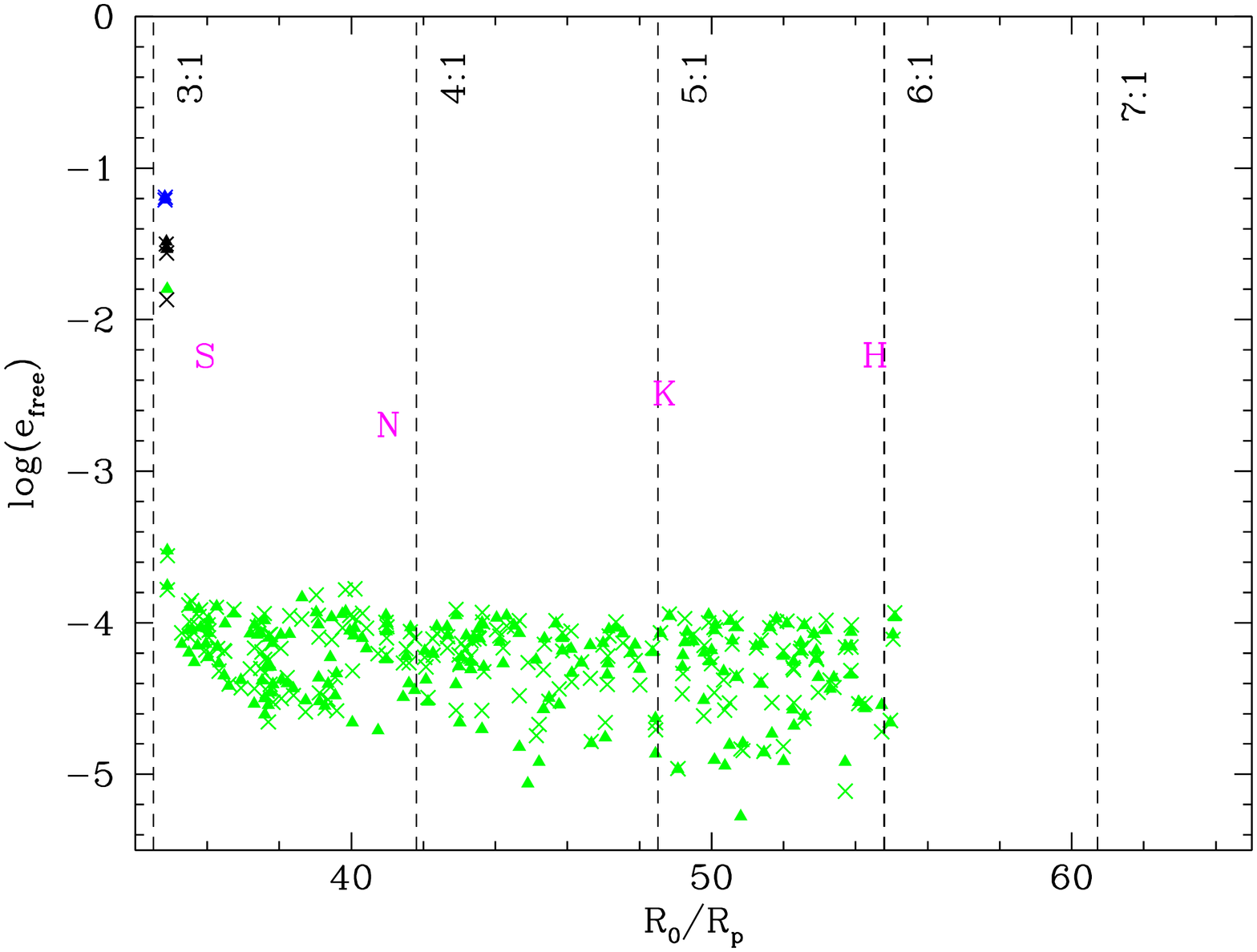}
\caption{
Same as Figures \ref{fig7} and \ref{fig8}, but for the large $A$ and initial $e_{\rm C} = 0$ integrations.
\label{fig9}
}
\end{center}
\end{figure}

The different results obtained with constant $\Delta t$ and constant
$Q$ may be due to the difference in tidal evolution timescale. As
mentioned, the constant $Q$ model takes around 5 to 10 times longer to
evolve. To test this, we increase the evolution timescale of the constant
$\Delta t$ model by decreasing $\Delta t$ to half (300 s) or a quarter
(150 s) of its original value, but keeping $A$ constant. Figure
\ref{fig8.1} shows a plot similar to Figure \ref{fig8}, but including
the results for $0.5 \Delta t$ (red squares) and $0.25 \Delta t$ (blue
squares). We find that for a longer evolution timescale, test
particles which are unaffected by resonances survive with slightly
lower final $e_{\rm free}$ (i.e., slightly closer to $e_{\rm free}$
obtained for constant $Q$). This shows that increasing the tidal
evolution timescale can decrease the final $e_{\rm free}$ of the
surviving test particles. However, we only tested a subset of 25 test
particles and did not try even smaller $\Delta t$, and we cannot make a
concrete conclusion on whether the constant $\Delta t$ model can also
reproduce the eccentricities of the current satellites if $\Delta t$
is sufficiently small.

\subsection{Large $A$ and Initial $e_{\rm C} = 0$}

No test particles are ejected when initial $e_{\rm C}$ is set to 0 (see
Table \ref{table2}).
The strength of the resonant terms depend on the eccentricities of
both Charon and the test particle.
Since $e_{\rm C}$ stays at zero and we have damped down
$e_{\rm free}$ of the test particles initially,
the test particles are very unlikely to be affected by high-order
resonances, even as low as 4:1 or 5:1.
Therefore, the eccentricities of the test particles are not easily
forced up to high values due to trapping in MMR with Charon, and their
probability of survival is much higher than in the previous two cases.

In Figure \ref{fig9}, $\sim 98\%$ of the test particles are in green,
which indicates that most of the test particles are not affected by
resonances. Their final $e_{\rm free}$ are mostly within the range of
$10^{-5}$ to $10^{-4}$, which are at least an order of magnitude lower
than the eccentricities of the current four small satellites. For the
test particles not in green, they are either trapped in resonance or
affected by resonances but not trapped. They all end near the 3:1 MMR
with Charon. Their $e_{\rm free}$ are all within the range of 0.01 to
0.1, which are more than twice the eccentricities of the current four
small satellites. Compared to Styx, the test particles affected by resonances are located closer to the 3:1 MMR with Charon. Hence, we cannot find any test particles that match any one of the four small satellites in this case.

\section{SUMMARY AND DISCUSSION}
\label{sec:conclusions}

We have investigated the early in-situ formation scenario, which
suggests that the four small satellites of Pluto formed in a debris
disk near their current locations before the tidal evolution of
Charon, by using $N$-body simulations to study the effects of the
tidal evolution on the small satellites. The small satellites were
treated as test particles that are initially collisionally damped to
their coldest orbits, and the system was integrated in two different
tidal models --- constant $\Delta t$ and constant $Q$, with different
relative rate of tidal dissipation in Charon and Pluto, $A$, and
initial $e_{\rm C}$. The plausibility of the early in-situ formation
scenario was assessed by comparing the final $R_0$ and $e_{\rm free}$
of the test particles with the actual values of the small satellites.

For large $A$ and initial $e_{\rm C} = 0$, all of the test particles
survive to the end. Most of the test particles are not affected by
resonances and their final $e_{\rm free}$ are at least an order of
magnitude lower than those of the current satellites. Test particles
that are affected by resonances are located closer to the 3:1
resonance than Styx, and their final $e_{\rm free}$ are at least twice
those of the current small satellites. Hence, we could not find any
test particles with orbits similar to the current small satellites in
this case.
For small $A$ and initial $e_{\rm C} = 0.2$, most of the test
particles are ejected, and nothing can be found closer than the 5:1
resonance. Hence this situation cannot explain the formation of Styx
and Nix which are now located closer than the 4:1 resonance.
For large $A$ and initial $e_{\rm C} = 0.2$, the results are different
for the two tidal models. For constant $\Delta t$, we found that
nothing matches the current satellites. For constant $Q$, we found 
that some test particles not affected by MMR survive with orbits
similar to the orbits of the four small satellites. However, there is
no preference for near resonance locations for these test particles.
We also tested the relation between tidal evolution timescale and
final $e_{\rm free}$ for the test particles that are not affected by
MMR by changing the $\Delta t$ value in the constant $\Delta t$
model. We discovered that increasing the tidal evolution timescale
slightly decreases the final $e_{\rm free}$ of the surviving test
particles, but a more complete set of simulation is needed to
determine whether the different results for the constant $\Delta t$
and constant $Q$ models are primarily due to the difference in the
tidal evolution timescale.

To conclude, the only case with test particles that survive to the end
of tidal evolution with similar orbits as the current four small
satellites is constant $Q$ with large $A$ and initial $e_{\rm C} =
0.2$.
However, we still need to explain the near resonance locations of the
small satellites for the early in-situ formation scenario to work.
Since the probability of randomly forming the small satellites near
MMR is low, there should be a reason for them to be near resonances,
and a successful satellite formation model needs to account for this
orbital feature.

We have assumed that the orbits of the small satellites are coplanar
with that of Charon in our study.
\cite{qui17} have recently shown that the high obliquities of the
small satellites (in particular, Styx and Nix) could be caused by
commensurability between the MMR frequency and spin precession rate if
the small satellites were captured into MMR involving inclination.
It is unclear whether this mechanism could work in the context of the
early in-situ formation scenario, as their simulations show that the
eccentricity of Nix is also excited to values much higher than the
observed eccentricity, because the lowest-order 4:1 resonant angles
containing the longitude of the ascending node $\Omega$ (which excite
the inclination of Nix) also involve the longitude of periapse
$\varpi$ (which excite the eccentricity of Nix).
However, their simulations assume initial $e_c = 0$.
If $e_c$ is nonzero at the time of the resonance capture, it may be
possible to keep the eccentricity of Nix small, as there are resonant
angles involving the longitude of periapse of Charon $\varpi_{\rm C}$,
but not $\varpi$.

We have neglected the masses of the small satellites in our study,
and the masses of especially the more massive Nix and Hydra are an
important factor in determining the long-term stability of the
satellites.
In particular, we may not need to explain the near resonance
location of Kerberos, since it is located in the only stable region
between Nix and Hydra \citep{pir11,you12}.
\cite{you12} found that the masses of Nix and Hydra should be smaller
than $5 \times 10^{16}\,$kg and $9 \times 10^{16}\,$kg, respectively,
in order for Kerberos to be stable over the age of the solar system.

As mentioned, some of the other formation scenarios for the small
satellites have been proven to be unsuccessful.
For example, the collisional capture scenario, in which the small
satellites were formed at their current locations from a collision
between two planetesimals captured by Pluto-Charon after tidal
evolution, was ruled out by \cite{pir12}, because they found that the
timescale of temporary capture for a planetesimal that is massive
enough to produce Nix and Hydra is much shorter than the timescale for
another planetesimal to come in and collide with it.
Their assumed masses for Nix and Hydra are based on the values reported
by \cite{tho08}, which are $5.8 \times 10^{17}\,$kg and $3.2 \times
10^{17}\,$kg, respectively.
These are much higher than the nominal masses in the latest measurements
by \cite{sho15} (see Table \ref{table0}) and the mass constraints for
Nix and Hydra in \cite{you12}.
If we study the collisional capture scenario by adopting the smaller
masses from the latest measurement, a smaller difference in the
timescales should be obtained. 

Since the masses of Nix and Hydra are an important factor
constraining the formation of the small satellites, we estimate
their masses from the latest size measurements by the {\it New
  Horizons} spacecraft.
Assuming that both are approximately ellipsoids, the size of Nix is
$50 \times 35 \times 33\,$km ($\pm 3\,$km), and the size of Hydra is
$65 \times 45 \times 25\,$km ($\pm 10\,$km) \citep{wea16}.
If we assume that
they are pure icy objects with density $\rho = 1\,{\rm g}\,{\rm
  cm}^{-3}$, the mass of Nix is $1.89 \times 10^{17}$ to $3.04 \times
10^{17}\,$kg, and the mass of Hydra is $1.21 \times 10^{17}$ to $6.05
\times 10^{17}\,$kg ($\pm 1\sigma$ from size). Even the lower masses
of Nix and Hydra we just estimated are more than $1 \sigma$ above
the upper limits measured by \cite{sho15} (see Table
\ref{table0}).
Besides, the density we assumed is lower than the density of both
Pluto ($\rho_{\rm p} = 1.86\,{\rm g}\,{\rm cm}^{-3}$) and Charon
($\rho_{\rm c} = 1.70\,{\rm g}\,{\rm cm}^{-3}$).
The same timescale problem arises in the collisional capture scenario
based on our estimated values, since our values are comparable to
those of \cite{tho08}.
On the other hand, our estimation for the masses is much larger than
the upper limits for the masses of Nix and Hydra in order for Kerberos
to stay at its current orbit for the age of the solar system,
according to \cite{you12}.
We have to account for the stability of Kerberos, unless the Pluto
satellite system was formed much later than the solar system.
The high albedo of the small satellites (see Table \ref{table0}) may
be evidence for their late formation, but the densities of craters on
Nix and Hydra suggest surface ages of at least 4 billion years
\citep{wea16}.
A more precise measurement of
the orbits of the small satellites by the {\it New Horizons}
spacecraft and the {\it Hubble Space Telescope} may help us to accurately determine the masses of the satellites and hence understand the origin of the small satellites.

\acknowledgments
We thank Mark Showalter for providing the initial state vectors of
Pluto's satellites from Showalter \& Hamilton.
This work was supported by a postgraduate studentship at the
University of Hong Kong (M.Y.W.) and Hong Kong RGC grant HKU 7030/11P
(M.Y.W. and M.H.L.).

\clearpage
\begin{deluxetable}{lcccc}
\tablewidth{0pt}
\tablecaption{Orbital and Physical Parameters of the Four Small Satellites of Pluto
\label{table0}}
\tablehead{
\colhead{} & \colhead{Styx} & \colhead{Nix} & \colhead{Kerberos} & \colhead{Hydra}
}
\startdata
Semimajor axis $a$ (km) 			& 42656 & 48694 & 57783 & 64738\\
Eccentricity $e$ ($10^{-3}$) 	& 5.787  & 2.036	 & 3.280  & 5.862 \\
Period $P$ (days)		& 20.16155 & 24.85463 & 32.16756 & 38.20177\\
$P/P_{\rm C}$	& 3.156542 & 3.891302 & 5.036233 & 5.980963\\
$GM$ ($10^{-3}\,{\rm km}^3\,{\rm s}^{-2}$) & $0.0+1.0$ & $3.0\pm 2.7$ & $1.1 \pm 0.6$ & $3.2 \pm 2.8$\\
Size (km)		& $16 \times 9 \times8$ & $50 \times 35
\times 33$ & $19 \times 10 \times 9$ & $65 \times 45 \times25$\\
Geometric albedo	& $0.65 \pm 0.07$ & $0.56 \pm 0.05$ & $0.56 \pm 0.05$ & $0.83 \pm 0.08$\\
\enddata
\tablecomments{$P_{\rm C}$ is the orbital period of Charon.
Orbital parameters and $GM$ are from \cite{sho15}, and size and geometric albedo are from \cite{wea16}.}

\end{deluxetable}

\begin{deluxetable}{ll}
\tablewidth{0pt}
\tablecaption{Combination of $A$ and Initial $e_{\rm C}$ in Different Tidal Models
\label{table1}}
\tablehead{
\colhead{Constant $\Delta t$} & \colhead{Constant $Q$}
}
\startdata
$e_{\rm C}=0$, $A=40$ & $e_{\rm C}=0$, $A=2.5$\\
$e_{\rm C}=0.2$, $A=10$ & $e_{\rm C}=0.2$, $A=0.65$\\
$e_{\rm C}=0.2$, $A=40$ & $e_{\rm C}=0.2$, $A=2.5$\\
\enddata
\end{deluxetable}

\clearpage
\begin{deluxetable}{lccccc}
\tablewidth{0pt}
\tablecaption{Statistics of Test Particles with Different Evolution
\label{table2}}
\tablehead{
\colhead{} & \colhead{Ejected} & \colhead{Still Trapped} & \colhead{Once Trapped} & \colhead{Affected} & \colhead{Unaffected}
\\
\colhead{} & \colhead{} & \colhead{in MMR} & \colhead{in MMR} & \colhead{by MMR} & \colhead{by MMR}
}
\startdata
\multicolumn{6}{c}{Small $A$ and Initial $e_{\rm C} = 0.2$} \\
\noalign{\vskip .7ex}
\cline{1-6}
\noalign{\vskip .7ex}
Constant $\Delta t$ 	& 160	& 4	& 6 		& 30 	& 0 \\
Constant $Q$		& 190	& 1 	& 4 		& 5 		& 0 \\
\noalign{\vskip .7ex}
\cline{1-6}
\noalign{\vskip .7ex}
\multicolumn{6}{c}{Large $A$ and Initial $e_{\rm C} = 0.2$} \\
\noalign{\vskip .7ex}
\cline{1-6}
\noalign{\vskip .7ex}
Constant $\Delta t$ 	& 56	& 47	& 0 		& 1 		& 96 \\
Constant $Q$		& 41	& 16 	& 1 		& 16 	& 126 \\
\noalign{\vskip .7ex}
\cline{1-6}
\noalign{\vskip .7ex}
\multicolumn{6}{c}{Large $A$ and Initial $e_{\rm C} = 0$} \\
\noalign{\vskip .7ex}
\cline{1-6}
\noalign{\vskip .7ex}
Constant $\Delta t$ 	& 0	& 2	& 0 		& 2 		& 196 \\
Constant $Q$		& 0	& 2 	& 0 		& 2 		& 196 \\
\enddata
\tablecomments{See Section \ref{sec:results} for the definition of each type of evolution.}
\end{deluxetable}

\clearpage

\clearpage

\clearpage

\clearpage

\clearpage

\clearpage

\clearpage

\clearpage

\clearpage

\clearpage

\clearpage

\begin{thebibliography}{}
\bibitem[Bromley \& Kenyon(2015)]{brom15}
     Bromley, B. C., \& Kenyon, S. J. 2015, ApJ, 809, 88
\bibitem[Brozovi\'c et al.(2015)]{bro15}
     Brozovi\'c, M., Showalter, M. R., Jacobson, R. A., \& Buie, M. W. 2015, Icarus, 246, 317
\bibitem[Canup(2005)]{can05}
     Canup, R. M. 2005, Science, 307, 546
\bibitem[Cheng et al.(2014a)]{che14a}
     Cheng, W. H., Lee, M. H., \& Peale, S. J. 2014a, Icarus, 233, 242
\bibitem[Cheng et al.(2014b)]{che14b}
     Cheng, W. H., Peale, S. J., \& Lee, M. H. 2014b, Icarus, 241, 180
\bibitem[Christy \& Harrington(1978)]{chr78}
     Christy, J. W., \& Harrington, R. S. 1978, AJ, 83, 1005
\bibitem[Holman \& Wiegert(1999)]{hol99}
     Holman, M. J., \& Wiegert, P. A. 1999, AJ, 117, 621
\bibitem[Kenyon \& Bromley(2014)]{ken14}
     Kenyon, S. J., \& Bromley, B. C. 2014, AJ, 147, 8
\bibitem[Lee \& Peale(2002)]{lee02}
     Lee, M. H., \& Peale, S. J. 2002, ApJ, 567, 596
\bibitem[Lee \& Peale(2003)]{lee03}
     Lee, M. H., \& Peale, S. J. 2003, ApJ, 592, 1201
\bibitem[Lee \& Peale(2006)]{lee06}
     Lee, M. H., \& Peale, S. J. 2006, Icarus, 184, 573
\bibitem[Leung \& Lee(2013)]{leu13}
     Leung, G. C. K., \& Lee, M. H. 2013, AJ, 763, 107
\bibitem[Levison \& Duncan(1994)]{lev94}
     Levison, H. F., \& Duncan, M. J. 1994, Icarus, 108, 18
\bibitem[Lithwick \& Wu(2008)]{lit08}
     Lithwick, Y., \& Wu, Y. 2008, preprint (arXiv:0802.2951)
\bibitem[Mignard(1980)]{mig80}
     Mignard, F. 1980, Moon Planets, 23, 185
\bibitem[Murray \& Dermott(1999)]{mur99}
     Murray, C. D., \& Dermott, S. F. 1999, Solar System Dynamics (Cambridge: Cambridge Univ. Press)
\bibitem[Pires dos Santos et al.(2011)]{pir11}
     Pires dos Santos, P. M., Giuliatti Winter, S. M., \& Sfair, R. 2011, MNRAS, 410, 273
\bibitem[Pires dos Santos et al.(2012)]{pir12}
     Pires dos Santos, P. M., Morbidelli, A., \& Nesvorn\'y, D. 2012,
     Celest.\ Mech.\ Dyn.\ Astron., 114, 341
\bibitem[Quillen et al.(2017)]{qui17}
     Quillen, A. C., Nichols-Fleming, F., Chen, Y.-Y., \& Noyelles, B. 2017, Icarus, 293, 94
\bibitem[Showalter \& Hamilton(2015)]{sho15}
     Showalter, M. R., \& Hamilton, D. P. 2015, Nature, 522, 7554
\bibitem[Showalter et al.(2011)]{sho11}
     Showalter, M. R., Hamilton, D. P., Stern, S. A., Weaver, H. A., Steffl, A. J., \& Young, L. A. 2011, IAU Circ.\ 9221
\bibitem[Showalter et al.(2012)]{sho12}
     Showalter, M. R., Weaver, H. A., Stern, S. A., Steffl, A. J., Buie, M. W., Merline, W. J., Mutchler, M. J., Soummer, R., \& Throop, H. B. 2012, IAU Circ.\ 9253
\bibitem[Smullen \& Kratter(2017)]{smu17}
     Smullen, R. A., \& Kratter, K. M. 2017, MNRAS, 466, 4480
\bibitem[Stern et al.(2015)]{ste15}
     Stern, S. A., et al. 2015, Science, 350, 6258
\bibitem[Tholen et al.(2008)]{tho08}
     Tholen, D. J., Buie, M. W., \& Grundy W. M. 2008, AJ, 135, 3
\bibitem[Walsh \& Levison(2015)]{wal15}
     Walsh, K. J., \& Levison, H. F. 2015, AJ, 150, 1
\bibitem[Ward \& Canup(2006)]{war06}
     Ward, W. R., \& Canup, R. M. 2006, Science, 313, 5790
\bibitem[Weaver et al.(2006)]{wea06}
     Weaver, H. A., et al. 2006, Nature, 439, 7079
\bibitem[Weaver et al.(2016)]{wea16}
     Weaver, H. A., et al. 2016, Science, 351, 6279 
\bibitem[Wisdom \& Holman(1991)]{wis91}
     Wisdom, J., \& Holman, M. 1991, AJ, 102, 1528
\bibitem[Woo \& Lee(2017)]{woo17}
     Woo, M. Y., \& Lee, M. H. 2017, AJ, submitted
\bibitem[Yoder \& Peale(1981)]{yod81}
     Yoder, C. F., \& Peale, S. J. 1981, Icarus, 47, 1
\bibitem[Youdin et al.(2012)]{you12}
     Youdin, A. N., Kratter K. M., \& Kenyon, S. J. 2012, ApJ, 755, 17
     
     
\end{thebibliography}
\end{document}